\documentclass[12pt]{iopart}
\usepackage{iopams}
\usepackage{bbold}
\usepackage{graphicx}
\usepackage{float}

\begin{document}
\title[]{Dissipative dynamics of atom-field entanglement in the ultrastrong-coupling regime}
\author[]{Ferdi Altintas}
\address{Department of Physics, Abant Izzet Baysal University, Bolu, 14280, Turkey}
\ead{ferdialtintas@ibu.edu.tr}
\begin{abstract}
The dynamics of atom-field entanglement for a system composed of two atoms resonantly coupled to a single mode leaky cavity field has been investigated beyond rotating wave approximation~(RWA). By using monogamic relation for entanglement of formation~(EOF) as well as the lower bound of EOF for bipartite mixed states in higher dimensions, contrary to the RWA case, the atom-field system in the steady states is found to be entangled in the strong-coupling regime and the entanglement can grow as a function of atom-field coupling strength.
\end{abstract}
\section{Introduction}
The light-matter interaction is one of the fundamental problems investigated in many aspects of modern physics ranging from quantum optics to quantum information processing and to condensed matter physics. The simple description of quantum light-matter interaction, namely the interaction of a qubit with a harmonic oscillator, is given by Rabi model~\cite{rabi36}. Indeed, this simple model has been recently used to describe many physical phenomena, such as an electron coupled to a phonon mode~\cite{holstein59}, an atom interacting with an electromagnetic field in a cavity~\cite{allen},  a superconducting qubit interacting with a nanomechanical resonator~\cite{irish03}, a transmission line resonator~\cite{blais04} or an LC circuit~\cite{chiorescu04}, etc~\cite{solano11}. Due to the excitation non-conserving nature of the Rabi model, it is hard to obtain an elegant formula for its eigen-spectrum; only recently Braak was given the spectrum in terms of a solution of a transcendental equation with a single variable~\cite{braak11}. For the weak atom-field coupling regime and the nearly resonant condition, it is legitimate to drop the excitation non-conserving terms, the so-called counter-rotating terms, in the atom-field interaction. This approximation is known as rotating wave approximation~(RWA) and has been well applied recently in most optical settings both experimentally and theoretically. Under RWA, Rabi model reduces to Jaynes-Cummings model~\cite{jaynes63} which is exactly solvable. On the other hand, with the recent technological advances in the circuit and cavity QED systems~\cite{irish03,blais04,chiorescu04,solano11}, it is now possible to engineer the strong atom-field couplings which makes crucial to consider the effects of counter-rotating terms on the dynamics of atom-field systems.

In the present study, we consider a model of two noninteracting qubits strongly coupled to a harmonic oscillator in a dissipative cavity. Our main focus is to investigate the effects of counter-rotating terms on the dissipative dynamics of entanglement between one of the atom and the cavity mode by using the recently derived Lindblad type quantum optical master equation~\cite{beaudoin11} which is adapted to the strong-coupling regime under Born and Markov approximations. The entanglement between atoms in the same setting has been recently investigated by many groups in the strong-coupling regime by using different master equation approaches~\cite{altintas13,altintas12,ficek10,jing09,jing08,ma12,wang13}. It has been shown that the entanglement between the atoms displays richer dynamical features in comparison to the RWA one. In particular, counter-rotating terms are found to lead to steady state entanglement between the atoms even for the initial states~(including vacuum state) that have no overlap with the decoherence free subspace of the system~\cite{altintas13}. Different from the previous studies~\cite{altintas13,altintas12,ficek10,jing09,jing08,ma12,wang13}, we focus on the entanglement for atom-field system in the strong-coupling regime. Under RWA conditions, even the entanglement between the atoms can reach a high non-zero value in the long-time limit for certain initial states, the cavity mode always becomes separable from the atoms in the steady states. By using monogamic relation for entanglement of formation~(EOF)~\cite{koashi04,fanchini11} as well as the lower bound of EOF for bipartite mixed states in higher dimensions~\cite{chen05}, contrary to the RWA case, we show that the atom-field system in the steady states can become entangled in the strong-coupling regime and the entanglement can grow as a function of atom-field coupling constant.
\section{The Model}\label{sec:model}
The Rabi Hamiltonian that describes the strong interaction of two identical qubits~(A and B) with a quantized cavity field F can be written as~($\hbar=1$)~\cite{rabi36}:
\begin{eqnarray}\label{rh}
H_R=\frac{1}{2}\omega_a\sum_{j=A,B}\sigma_z^j+\omega_f a^{\dagger}a+g\sum_{j=A,B}\sigma_x^j(a+a^{\dagger}),
\end{eqnarray}
where $\omega_a$ $(\omega_f)$ is the qubit transition (cavity field) frequency, $a$ $(a^{\dagger})$ is the annihilation (creation) operator of the cavity field, $\sigma_z$ $(\sigma_x)$ is the Pauli matrix in the $z$ $(x)$ direction and $g$ is the coupling constant between the atom and the cavity field. The terms, $\sigma_-^ja, \sigma_+^ja^{\dagger}$, in the interaction part of the Hamiltonian are called counter-rotating terms~(CRTs) and they do not conserve the excitation number; even the atoms and the field are initially in their ground states,  the time evolution of the mean excitation number is nonzero~\cite{altintas12,werlang08}. On the other hand, in the weak-coupling regime where $g<<\omega_f,\omega_a$, the effects of CRTs are negligible and these terms can be dropped from the Hamiltonian. This approximation is known as rotating wave approximation (RWA) which leads to two-qubit Jaynes-Cummings Hamiltonian~\cite{jaynes63}:
\begin{eqnarray}\label{jch}
H_{JC}=\frac{1}{2}\omega_a\sum_{j=A,B}\sigma_z^j+\omega_f a^{\dagger}a+g\sum_{j=A,B}\left[\sigma_+^ja+\sigma_-^ja^{\dagger}\right].
\end{eqnarray}

The photon loss due to the imperfections of the cavity mirrors should be taken into account. Under Born and Markov approximations~\footnote{The principal system comprised of atoms and the cavity field is assumed to weakly interact with an environment which is the source of non-unitary dynamics. The Born and Markov approximations are considered between the open system and environment that leads to Lindblad type master equation.}, the non-unitary dynamics of atoms-field system is generally represented by the "{\it standard} " Lindblad master equation:
\begin{eqnarray}\label{slme}
\dot{\rho}=-i[H_{JC},\rho]+\kappa D[a]\rho,
\end{eqnarray}
where $D[m]\rho=\frac{1}{2}\left(2m\rho m^{\dagger}-\rho m^{\dagger}m-m^{\dagger}m\rho\right)$, $\rho$ is the atoms-field density matrix and $\kappa$ is the photon leakage rate. In the weak atom-field coupling regime where RWA can be applied~(Eq.~(\ref{jch})), the master equation~(\ref{slme}) has been well studied and can be used accurately to describe many cavity and circuit QED systems. Contrary to the numerous investigations of the master equation~(\ref{slme}) in the strong atom-field coupling regime by using Rabi Hamiltonian~(\ref{rh}) in an {\it ad hoc} manner~(see Ref.~\cite{altintas13} and references therein),  it is, in fact, at the root of unphysical effects~\cite{beaudoin11}; it creates excess photons in atom-field system and produces spurious qubit flipping. The main reason of the unphysical predictions obtained from standard Lindblad master equation in the strong-coupling regime is to neglect the qubit-field interaction~($g\rightarrow 0$) while deriving the dissipator part of Eq.~(\ref{slme}). Although the error introduced by the approximation~($g\rightarrow 0$) is negligible under RWA, the qubit-field interaction becomes influential in the strong-coupling regime and causes the unphysical effects. Recently, a new Lindblad master equation, which carefully considers the transitions among the eigenstates of the Rabi Hamiltonian~(\ref{rh}), has been given~\cite{beaudoin11}. It can be written as~\cite{beaudoin11}:
\begin{eqnarray}\label{ilme}
\dot{\rho}=-i[H_R,\rho]+\displaystyle\sum_{j,k>j}\Gamma_{\kappa}^{jk}D\left[\left|j\right\rangle\left\langle k\right|\right]\rho,
\end{eqnarray}
where $\Gamma_{\kappa}^{jk}=\kappa\frac{\omega_k-\omega_j}{\omega_f}|\left\langle j\right|(a+a^{\dagger})\left|k\right\rangle |^2$ are the relaxation coefficients~\cite{ridolfo12}, $\left|j\right\rangle$ and $\omega_j$ define the spectrum of the Hamiltonian~(\ref{rh}) i.e., $H_R\left|j\right\rangle=\omega_j\left|j\right\rangle$ and the eigenstates should be labeled according to increasing energy~(label $\left|j\right\rangle$ such that $\omega_k>\omega_j$ for $k>j$). Also, note that as $g\rightarrow 0$, the standard dissipator~(\ref{slme}) can be recovered from Eq.~(\ref{ilme}). Here, we call the equation~(\ref{ilme}) as  the "{\it improved} " Lindblad master equation which, indeed, cures the unphysical effects originated from the use of the {\it ad hoc} master equation~(\ref{slme}) in the strong atom-field coupling regime~\cite{beaudoin11}.

For the numerical solution of the master equation~(\ref{ilme}) with Rabi model, we have considered basis vectors of type $\left|ij\right\rangle_{AB}\left|n\right\rangle_{F}\equiv\left|ijn\right\rangle$ where $i,j\in e,g$ are the excited and ground states of the qubit and $n=0,1,\ldots , d-1$. The field Fock space is truncated for large enough $d$ until the results converge~\cite{altintas13}. For the sake of simplicity, we will consider the resonant case, $\omega_f=\omega_a=\omega$, and the field initially prepared in its ground state $\left|0\right\rangle_{F}$.

Now, we investigate the dissipative dynamics of entanglement between one of the atom (say A) and the cavity mode F in the strong-coupling regime. The reduced density matrix of A-F system can be obtained by taking a partial trace of $\rho$ over the remaining part B: $\rho_{AF}=Tr_B\rho$. In practice, the considered system has a dimension of ($2\otimes\infty$). Since we apply a convergence criteria to the field Fock space to solve the master equation~(\ref{ilme}), the atom-field system becomes a qubit-qudit ($2\otimes d$) system. 
\section{Atom-field entanglement in the ultrastrong-coupling regime}\label{sec:results}
Entangled states are vital for many quantum computation and communication protocols. Entanglement of formation~(EOF) is one of the meaningful and physically motivated measure of entanglement. For two-qubit states, EOF can be calculated directly through the entanglement monotone, concurrence~\cite{wootters98}. For higher dimensional bipartite mixed states, except for some special cases~\footnote{One of the special case is the analytic calculation of EOF for a family of $(2\otimes d)$-dimensional bipartite mixed states reduced from tripartite ($2\otimes 2\otimes d$) pure states~\cite{koashi04,fanchini11} which will be discussed later in the text.}, the evaluation of EOF is a formidable task which can require highly complex optimization process. Therefore, recent efforts have led to several bounds to estimate the amount of entanglement in high dimensional quantum systems~\cite{chen05}. In the present study, we will use recently proposed the lower bound of EOF~\cite{chen05} to estimate the amount of entanglement for the atom-field states in the strong-coupling regime. Its definition is based on the comparison of two strong separability criteria: The positive partial transpose and the realignment criteria. It was shown that EOF is tightly bounded for qubit-qudit system as~\cite{chen05}:
\begin{equation}
E(\rho_{AF})\geq \left\{\begin{tabular}{ll} $0$ & if $\Lambda =1,$ \\ $H_{2}\left[\frac{1}{2}\left(1+\sqrt{1-(\Lambda-1)^2}\right)\right]$ & if $\Lambda \in \left[ 1, 2 \right],$\end{tabular}\right. \label{lbeof}
\end{equation}
where $\Lambda =\max (\left\Vert \rho_{AF} ^{T_{A}}\right\Vert,\left\Vert R(\rho_{AF})\right\Vert )$, $H_{2}(x)=-x\log_2(x)-(1-x)\log_2(1-x)$ and $\left\Vert G\right\Vert=Tr(GG^{\dagger })^{\frac{1}{2}}$ is the trace norm. The matrix $\rho_{AF}^{T_{A}}$ is the partial transpose with respect to the subsystem A. $\rho_{ij}^{T_i}\geq 0$ is, in fact, sufficient for separability of bipartite ($2\otimes 2$) and ($2\otimes 3$) states. $R(\rho_{AF})$ is the realigned version of $\rho_{AF}$~\cite{chen05} and is another operational criteria for separability which satisfies $\left\Vert R(\rho_{ij})\right\Vert\leq 1$ for any bipartite separable state $\rho_{ij}$.

We first investigate the dissipative dynamics of entanglement between the atom A and cavity mode F under RWA conditions. The dynamics of atoms-field system with cavity decay under RWA given by Eq.~(\ref{slme}) have been throughly investigated, recently~\cite{altintas13,mazzola09}. It was shown that the atoms-field steady states have a simple structure determined solely by the overlap of atomic initial state with the subradiant state~(see Refs.~\cite{altintas13,mazzola09} for details). More precisely, the cavity decay tends to drive the atoms-field system to the ground state~($\left|gg0\right\rangle$) of $H_{JC}$. On the other hand, the population of the subradiant state, $\left|\Phi^-\right\rangle=\frac{1}{\sqrt{2}}(\left|eg\right\rangle-\left|ge\right\rangle)$, given by $b=\left\langle \Phi^-\right|\rho_{AB}(0)\left|\Phi^-\right\rangle$ does not decay in the evolution process. Therefore, the atoms-field system is driven to a mixed state involving only the two decoherence free subspaces~(DFS): the ground state, $\left|gg0\right\rangle$, and the dark state, $\left|\Phi^-0\right\rangle=\frac{1}{\sqrt{2}}(\left|eg0\right\rangle-\left|ge0\right\rangle)$. Depending on the atomic initial state~\footnote{Here we consider initial states that have no initial coherence between ground and dark states, i.e., $\left\langle gg\right|\rho_{AB}(0)\left|\Phi^-\right\rangle=0$. Such states correspond to a wide range of initial states, including X states~\cite{altintas13}.}, the atoms-field steady states can be written as~\cite{altintas13}:
\begin{eqnarray}\label{afss}
\rho^{SS}=\left[(1-b)\left|gg\right\rangle\left\langle gg\right|+b\left|\Phi^-\right\rangle\left\langle\Phi^-\right|\right]\otimes\left|0\right\rangle\left\langle 0\right|.
\end{eqnarray}
One should note that the atoms-field steady states~(\ref{afss}) have very interesting features: {\bf (i)} A necessary condition for the long time asymptotic entanglement between the atoms is the non-zero overlap with the subradiant state, i.e., $b\neq 0$. {\bf (ii)} The atoms-field steady states only depends on the initial states, so they are independent of atom-field coupling strength, $g$. {\bf (iii)} The field mode always becomes separable from the atoms, even the atoms can reach a highly entangled states for $b\neq 0$. For instance, for an initial state $\left|\Psi(0)\right\rangle=\left|ge0\right\rangle$, we have in the steady states, $E(\rho_{AB})\approx 0.35$, while $E(\rho_{AF})=E(\rho_{BF})=0$. The breakdown of the descriptions {\bf (i)} and {\bf (ii)} under non-RWA conditions was analyzed in Ref.~\cite{altintas13} where the dissipative dynamics of entanglement between the atoms have been investigated in the strong-coupling regime. In the following, we will focus on the breakdown of the description {\bf (iii)} and show that the atom-field system can become entangled in the steady states under non-RWA conditions.
\begin{figure}[!ht]\centering
\includegraphics[width=7.7cm]{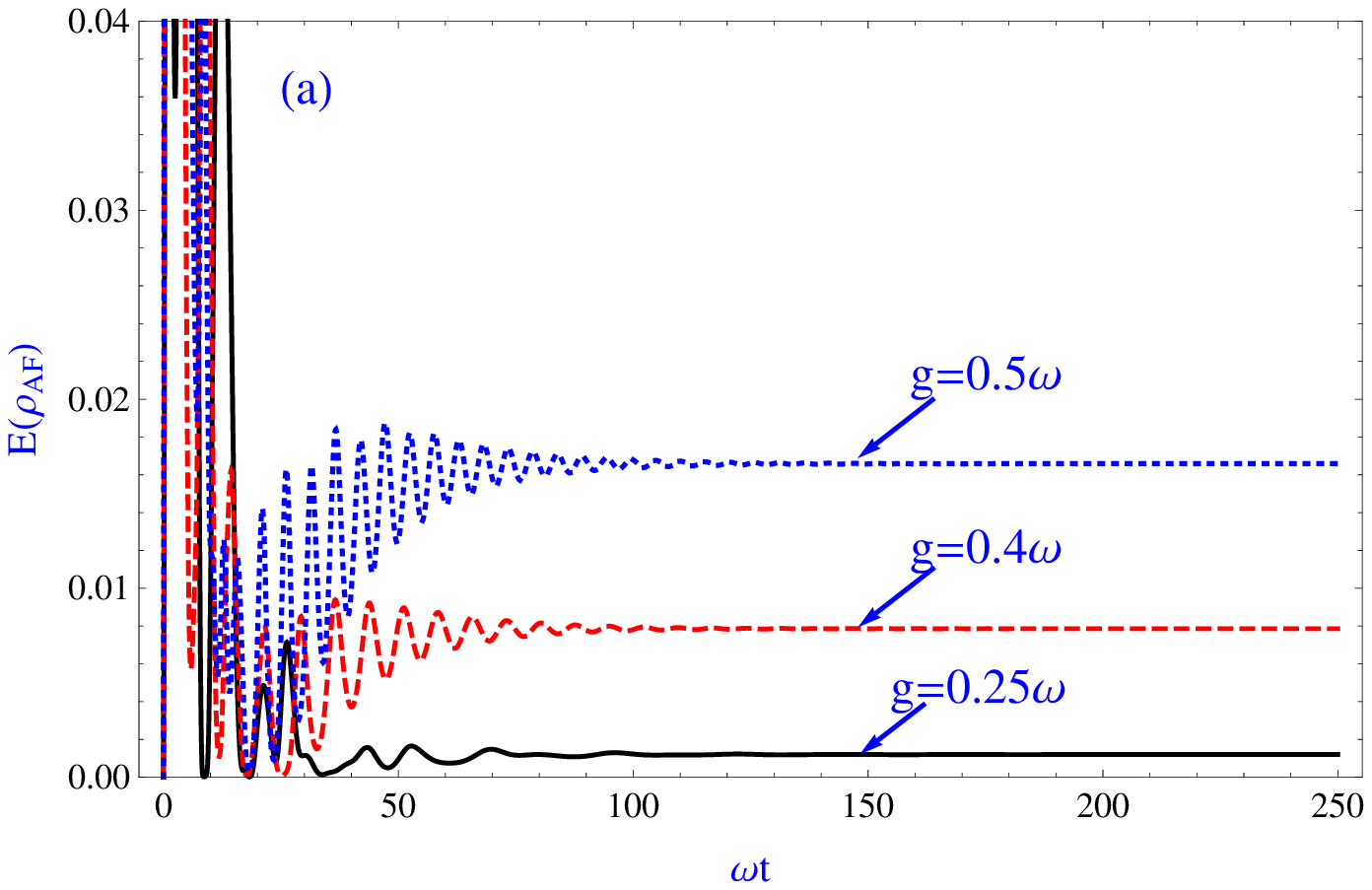}
\includegraphics[width=7.7cm]{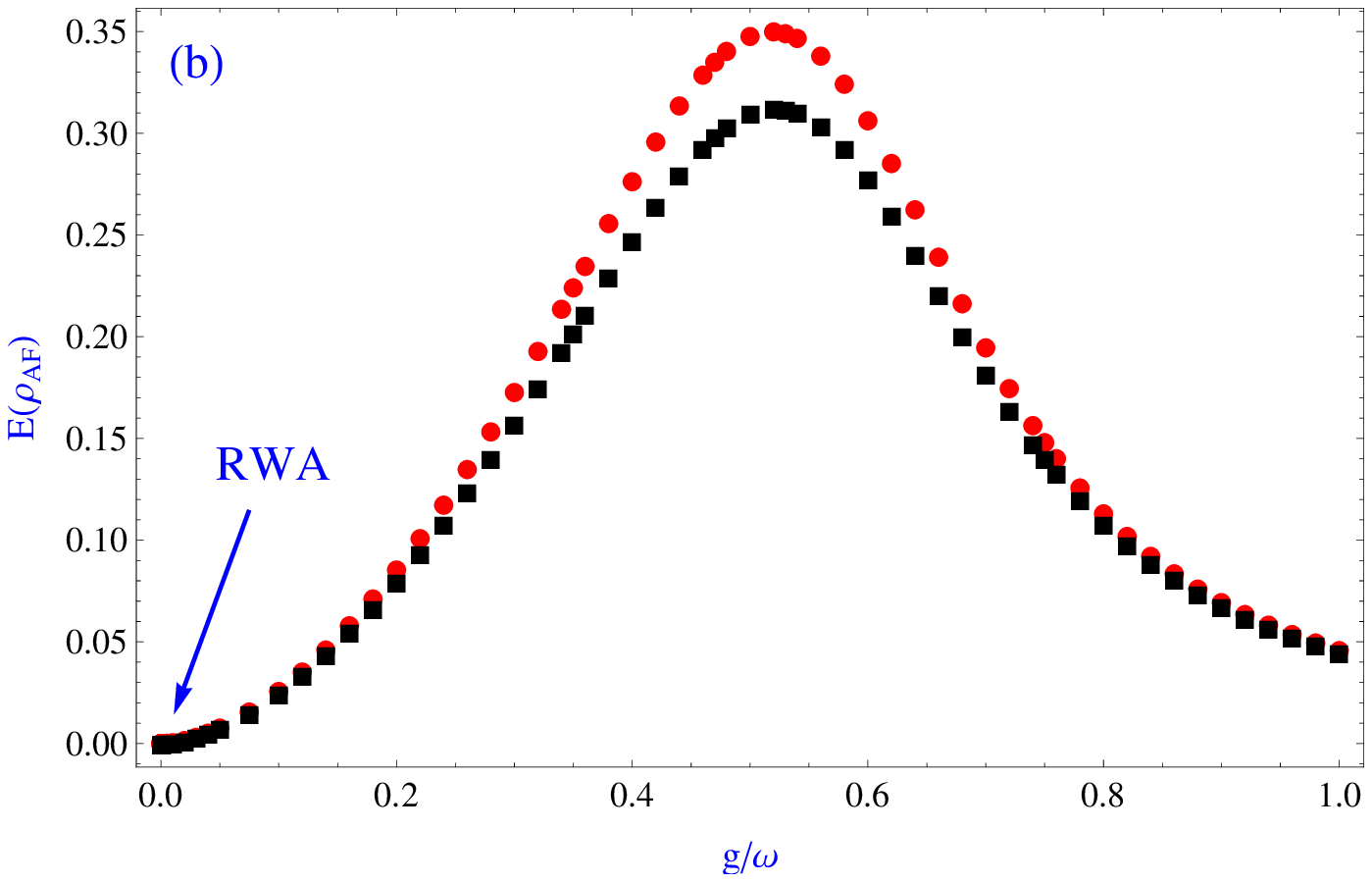}
\caption{\label{fig1} (a) Lower bound of EOF between the qubit A and the cavity mode F versus $\omega t$ for $\rho(0)=\left|ge0\right\rangle\left\langle ge0\right|$ in the strong-coupling regime with $g=0.25\omega$~(solid), $g=0.4\omega$~(dashed), $g=0.5\omega$~(dotted) and $\kappa=0.2\omega$ obtained by using the dissipator~(\ref{ilme}). (b) Ground state EOF between the qubit A and cavity mode F versus $g/\omega$ obtained from the monogamic relation~(circles) and from the lower bound~(rectangular).}
\end{figure}

In Fig.~\ref{fig1}, we investigate the dissipative dynamics of atom-field entanglement in the strong-coupling regime by using the "{\it improved} " dissipator, Eq.~(\ref{ilme}). Fig.~\ref{fig1}(a) shows the dynamics of lower bound of EOF~(Eq.~(\ref{lbeof})) for the initial state $\left|\Psi(0)\right\rangle=\left|ge0\right\rangle$ and for several $g/\omega$ values at $\kappa=0.2\omega$. It can be deduced from the figure that due to the interaction between the atom and field, entanglement can be induced in a short time even from an initial product state. However, the field does not become separable from the atoms in the asymptotic time limit in the strong-coupling regime. The simple example shown in Fig.~\ref{fig1}(a) demonstrates the breakdown of the above description {\bf (iii)} which states that $E(\rho_{AF})=0$ in the steady states under RWA. In fact, it was conjectured~\cite{altintas13} for the extended Werner-like atomic initial states that the atoms-field steady states under non-RWA dynamics have the same form of Eq.~(\ref{afss}) with just $\left|gg0\right\rangle$ replaced by the ground state, $\left|\widetilde{gg0}\right\rangle$, of the Rabi Hamiltonian~(\ref{rh}):
\begin{eqnarray}\label{afss2}
\rho^{SS}=(1-b)\left|\widetilde{gg0}\right\rangle\left\langle \widetilde{gg0}\right|+b\left|\Phi^-0\right\rangle\left\langle\Phi^-0\right|.
\end{eqnarray}
It is immediate to see from Eq.~(\ref{afss2}) that the atom-field entanglement, more precisely the role of CRTs, in the steady states stems from the ground state of Rabi Hamiltonian~\cite{ashhab10}. In Fig.~\ref{fig1}(b), we analyze the ground state entanglement for A-F system as a function of coupling strength. One should note that the ground state forms a tripartite pure state of dimensions $(2\otimes 2\otimes d)$. For a tripartite pure density matrix, EOF and quantum discord (or classical correlations~\cite{koashi04}) obey an important monogamic relation~\cite{fanchini11}: 
\begin{eqnarray}\label{eofmono}
E(\rho_{AF})=D(\rho_{AB})+S_{A|B},
\end{eqnarray}
where $S_{A|B}=S(\rho_{AB})-S(\rho_B)$~($S(\rho)$ is the von Neumann entropy) and $D(\rho_{AB})$ is the quantum discord between the atoms which can be calculated analytically for the ground state~\cite{altintas13,lee13}. Indeed, Eq.~(\ref{eofmono}) provides the exact calculation of EOF for a special class of higher dimensional bipartite mixed states. The circle line in Fig.~\ref{fig1}(b) displays the ground state EOF as a function of coupling strength $g/\omega$  obtained from monogamic relation, Eq.~(\ref{eofmono}). EOF grows from zero~(RWA regime) with the coupling up to $g/\omega\approx 0.5$ and drops as $g/\omega$ further increases after it approaches the maximum. In Fig.~\ref{fig1}(b), we also display the lower bound of EOF~(rectangular line) between A and F for the ground state and it is found to be in agreement with the exact EOF~\cite{lastra12}. One should note that for $g/\omega>0.5$, the number of virtually created photons increases significantly in the ground state as coupling increases. It seems that those virtual photons partially destroy the entanglement between the qubits~\cite{altintas13,lee13} and between the atom and cavity mode as shown in Fig~\ref{fig1}(b).
\section{Conclusions}\label{sec:conc}
The non-RWA dynamics of atom-field entanglement has been investigated for a system composed of two identical qubits resonantly coupled to a single mode leaky cavity field. The strong atom-field interaction is found to induce atom-field entanglement in the steady states contrary to the weak-coupling one where RWA is valid.

\end{document}